\documentclass{vgtc}                         
\usepackage{times}                    
         
\usepackage{tikz}
\usepackage{booktabs}
\usepackage{longtable} 
\usepackage{pgfplots}
\pgfplotsset{compat=1.17}
\usepackage{pgfplotstable}
\usepackage[table]{xcolor}
\usetikzlibrary{shapes, arrows.meta, positioning, decorations.pathreplacing, calc}
\usepackage{graphicx}

\usepackage{mathptmx}                  
\onlineid{1292}

\vgtccategory{Training and Education}

\vgtcinsertpkg

\title{XR Design Framework for Early Childhood Education}

\author{Supriya Khadka\thanks{e-mail: khadkas25@uni.coventry.ac.uk}\\ %
        \scriptsize Coventry University %
\and Sanchari Das\thanks{e-mail: sdas35@gmu.edu}\\ %
     \scriptsize George Mason University %
     }

\abstract{
    Extended Reality in early childhood education presents high-risk challenges due to children's rapid developmental changes. While augmented and virtual reality offer immersive pedagogical benefits, they often impose excessive cognitive load or sensory conflict. We introduce the \textbf{Augmented Human Development (AHD)} framework to model these interactions through cognitive, sensory, environmental, and developmental parameters. To ground this framework, we conducted a Systematization of Knowledge (SoK) of $111$ peer-reviewed studies involving children aged $3-8$. Our findings, interpreted through the AHD lens, reveal a critical ``risk vs. attention gap,'' where high-impact safety and security risks remain under-researched compared to short-term pedagogical gains.

} 

\begin{document}

\firstsection{Introduction}

\maketitle

Early childhood is one of the most formative periods of human development, during which children rapidly acquire cognitive, emotional, social, and motor competencies that shape later behavior~\cite{unesco_ecce_need_know_2025}. Extended Reality (XR) is an umbrella term encompassing Augmented Reality (AR), which overlays virtual elements on the physical world; Virtual Reality (VR), which replaces the environment via head-mounted displays (HMDs); and Mixed Reality (MR), which supports spatial mapping and bidirectional interaction~\cite{fast2018testing}. Although XR can increase engagement in domains like STEM and literacy~\cite{noah2021exploring}, its use in early childhood education (ECE) is high risk because systems calibrated for adults can create sensory conflict or exceed developmental limits~\cite{bexson2024safety}. Despite these concerns, there is a lack of structured frameworks that link XR pipeline properties to cognitive load and access inequity. This work addresses this gap via the Augmented Human Development (AHD) framework, grounded in a Systematization of Knowledge (SoK) of $111$ studies identifying misalignment with child development.

\section{Augmented Human Development Framework}
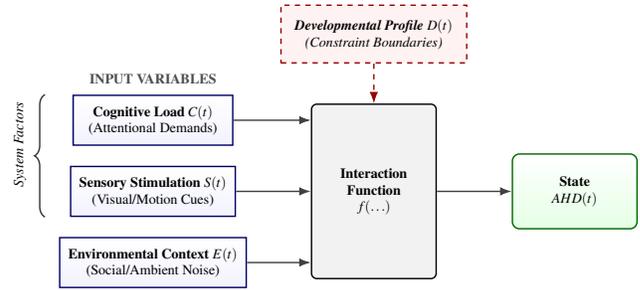
\begin{figure}[htbp]
    \centering
    \resizebox{\linewidth}{!}{
    \begin{tikzpicture}[
        node distance=0.8cm and 1.5cm, 
        inputblock/.style={
            rectangle, 
            draw=blue!40!black, 
            top color=white, 
            bottom color=blue!5, 
            thick, 
            inner sep=5pt,
            minimum width=3.2cm, 
            minimum height=1cm, 
            align=center, 
            font=\scriptsize
        },
        processblock/.style={
            rectangle, 
            draw=black, 
            fill=gray!10, 
            thick, 
            rounded corners=3pt,
            minimum width=2.5cm, 
            minimum height=3.5cm, 
            align=center, 
            font=\footnotesize\bfseries
        },
        constraintblock/.style={
            rectangle, 
            draw=red!60!black, 
            dashed, 
            inner sep=8pt,
            thick, 
            fill=red!5, 
            minimum width=3.5cm, 
            minimum height=0.8cm, 
            align=center, 
            font=\scriptsize\itshape
        },
        outputblock/.style={
            rectangle, 
            draw=green!40!black, 
            top color=white, 
            bottom color=green!10, 
            thick, 
            rounded corners, 
            minimum width=2.5cm, 
            minimum height=1.5cm, 
            align=center, 
            font=\footnotesize\bfseries
        },
        line/.style={draw, -Latex, thick, gray!50!black}
    ]

    \node[inputblock] (C) {\textbf{Cognitive Load} $C(t)$ \\ \scriptsize (Attentional Demands)};
    \node[inputblock, below=0.4cm of C] (S) {\textbf{Sensory Stimulation} $S(t)$ \\ \scriptsize (Visual/Motion Cues)};
    \node[inputblock, below=0.4cm of S] (E) {\textbf{Environmental Context} $E(t)$ \\ \scriptsize (Social/Ambient Noise)};

    \node[above=0.1cm of C, font=\bfseries\footnotesize, color=gray!60!black] {INPUT VARIABLES};

    \node[processblock, right=1.5cm of S] (Func) {Interaction\\Function\\$f(\dots)$};

    \node[constraintblock, above=0.8cm of Func] (D) {\textbf{Developmental Profile} $D(t)$ \\ \scriptsize (Constraint Boundaries)};

    \node[outputblock, right=1.5cm of Func] (Out) {State\\$AHD(t)$};

    \draw[line] (C.east) -- (C.east -| Func.west);
    \draw[line] (S.east) -- (Func.west);
    \draw[line] (E.east) -- (E.east -| Func.west);
    \draw[line, dashed, red!60!black] (D.south) -- (Func.north);
    \draw[line] (Func.east) -- (Out.west);

    \draw[decorate, decoration={brace, amplitude=6pt}, thick, gray!40!black]
        ([xshift=-0.5cm]S.south west) -- ([xshift=-0.5cm]S.south west |- C.north west)
        node[midway, left=0.5cm, yshift=1.2cm, rotate=90, font=\scriptsize\itshape, color=black] {System Factors};

    \end{tikzpicture}
    }
    
    \caption{Conceptual Definition of the AHD Framework}
    \label{fig:ahd-framework}
\end{figure}

The core contribution of this work is the AHD framework (Figure~\ref{fig:ahd-framework}), which frames XR as a dynamic modulator of developmental processes. At any moment \(t\), child-system interaction is defined as:

\begin{equation}
AHD(t) = f\big(C(t), S(t), E(t), D(t)\big),
\end{equation}

where \(C(t)\) \textbf{(Cognitive Load)}: Denotes demands on attention, memory, and executive functions; overload occurs when children focus on interface mechanics over learning goals, \(S(t)\) \textbf{(Sensory Stimulation)}: Captures motion cues, visual complexity, and sound; unstable augmentations disrupt this dimension, causing disorientation, \(E(t)\) \textbf{(Environmental Context)}: Reflects teacher mediation, social collaboration, and ambient noise, including instructional readiness and classroom social ecology, and \(D(t)\) \textbf{(Developmental Profile)}: Accounts for age, motor abilities, and sensory thresholds, defining strict boundaries for acceptable workloads.

The AHD framework demonstrates that common XR challenges reported in the literature, such as cognitive overload, novelty-driven distraction, VR-induced fatigue, or inequitable access, are not isolated failures but coupled effects that emerge from misalignment among \(C(t)\), \(S(t)\), \(E(t)\), and \(D(t)\).

\section{Method: Systematization of Knowledge (SoK)}

To ground the AHD framework, we conducted a systematic review of $111$ peer-reviewed studies ($2010-2025$) involving participants aged $3-8$, filtered from 2,198 records across eight databases (ACM Digital Library, IEEE Xplore, Scopus, ERIC, ProQuest, Taylor \& Francis, Wiley Online Library, and EBSCOhost). For eligible studies, we extracted metadata, participants' information and feasibility details. The studies were scored on a $0-2$ scale (0 = not addressed, 1 = mentioned, 2 = substantively discussed) across seven dimensions. We define \textbf{Scholarly Attention} as the mean score within each dimension to quantify academic focus. The dimensions map to AHD parameters: \textbf{Pedagogy} (curriculum alignment), \textbf{Privacy} (consent and compliance),\textbf{ Data Security} (encryption and storage), \textbf{Medical/Health} (cybersickness and ergonomics), \textbf{Technical} (hardware latency), \textbf{Disability Access} (special needs support), and \textbf{Low-Resource Access} (low-cost device suitability). We also independently calculated a \textbf{Real-World Risk} score (Likelihood $\times$ Impact) on a $1-9$ scale. Values were informed by external industry standards such as NIST and WHO to identify research gaps.

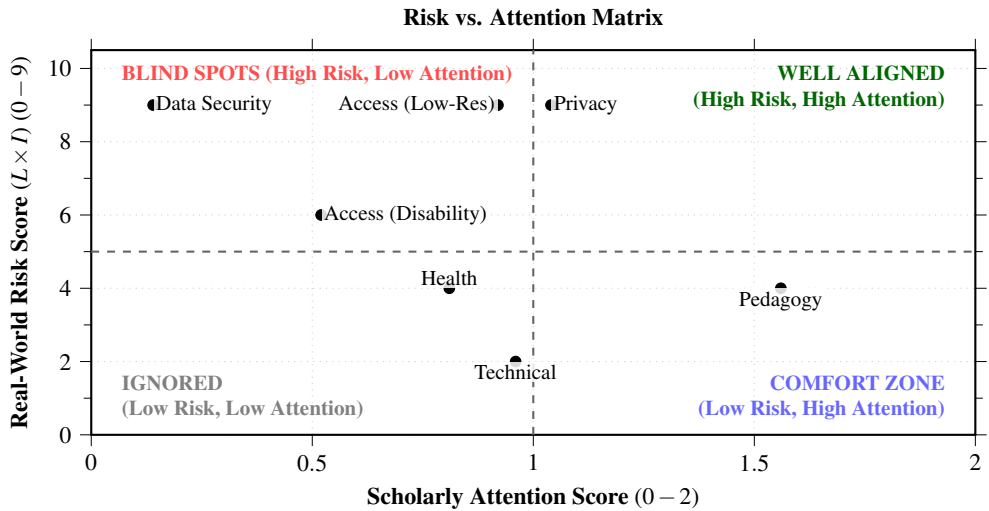
\begin{figure*}[htbp]
\centering
    \begin{tikzpicture}
        \begin{axis}[
            width=0.75\linewidth,
            height=6.7cm,
            xlabel={\textbf{Scholarly Attention Score} ($0-2$)},
            ylabel={\textbf{Real-World Risk Score} ($L \times I$) ($0-9$)},
            xmin=0, xmax=2,
            ymin=0, ymax=10.5,
            xtick={0, 0.5, 1, 1.5, 2},
            ytick={0, 2, 4, 6, 8, 10},
            minor y tick num=1,
            axis line style={draw=black, thick},
            tick style={draw=black},
            tick align=outside,
            grid=major,
            grid style={dotted, black!20},
            title={\textbf{Risk vs. Attention Matrix}},
            clip=false,
            scatter/classes={
                a={mark=*, blue!60, draw=black, mark size=3pt},
                b={mark=*, red!70, draw=black, mark size=3pt},
                c={mark=square*, orange, draw=black, mark size=3pt}
            }
        ]
            \draw[black!60, dashed, thick] (axis cs:1, 0) -- (axis cs:1, 10.5);
            \draw[black!60, dashed, thick] (axis cs:0, 5) -- (axis cs:2, 5);

            \node[anchor=north west, align=left, font=\bfseries\scriptsize, color=red!70] 
                at (axis cs:0.05, 10.3) {BLIND SPOTS (High Risk, Low Attention)};
            
            \node[anchor=north east, align=right, font=\bfseries\scriptsize, color=green!40!black] 
                at (axis cs:1.95, 10.3) {WELL ALIGNED\\(High Risk, High Attention)};
            
            \node[anchor=south east, align=right, font=\bfseries\scriptsize, color=blue!60] 
                at (axis cs:1.95, 0.2) {COMFORT ZONE\\(Low Risk, High Attention)};
            
            \node[anchor=south west, align=left, font=\bfseries\scriptsize, color=gray] 
                at (axis cs:0.05, 0.2) {IGNORED\\(Low Risk, Low Attention)};

            \addplot[
                scatter,
                only marks,
                scatter src=explicit symbolic,
                visualization depends on={value \thisrow{label}\as\mylabel},
                visualization depends on={value \thisrow{anchor}\as\myanchor},
                nodes near coords={\mylabel},
                every node near coord/.append style={anchor=\myanchor, font=\scriptsize, color=black, fill=white, inner sep=1pt, opacity=0.85, text opacity=1}
            ] table [meta=class] {
                x      y    class   label                 anchor
                0.14   9    b       {Data Security}       west
                0.52   6    b       {Access (Disability)} west
                1.04   9    c       Privacy               west
                0.92   9    c       {Access (Low-Res)}    east
                1.56   4    a       Pedagogy              north
                0.96   2    a       Technical             north
                0.81   4    a       Health                south
            };
        \end{axis}
    \end{tikzpicture}
   \caption{Risk Assessment Matrix Mapping Scholarly Attention Against Calculated Real-World Risk}
    \label{fig:risk-matrix}
\end{figure*}

\section{Results and Discussion}
Our analysis of the $111$ studies shows a strong emphasis on AR (73\%), followed by VR (20\%) and MR (4.5\%). Most deployments relied on mobile and tablet-based systems (63\%), while head-mounted displays (9.9\%) and specialized custom setups (13.5\%) appeared far less frequently.

\subsection{Systemic Frictions and Coupled Effects}
The AHD framework demonstrates that technical and pedagogical failures do not occur in isolation. Instead, they arise from \textbf{bio-technical mismatches} caused by friction between system requirements and the developing child's body. In \textbf{handheld AR}, studies report high rates of target-tracking failure because monocular camera sensing demands stability ($S(t)$) that exceeds the fine motor control capabilities of children aged $3$ to $8$ ($D(t)$). This mismatch produces visual jitter, frequent tracking loss, and interaction breakdowns. \textbf{Immersive VR} systems primarily fail due to hardware and anthropometric incompatibilities. Standard head-mounted displays do not account for early childhood cranial dimensions ($D(t)$), which leads to optical misalignment, occlusion, and musculoskeletal strain. These systems also introduce persistent neuro-vestibular conflict when system-generated visual motion cues ($S(t)$) diverge from the child's physical vestibular signals ($D(t)$), increasing the risk of cybersickness. Pedagogical integration faces comparable \textbf{instructional and ecological barriers}. Although 107 of the 111 studies report positive learning outcomes, many of these effects reflect short-term novelty rather than sustained conceptual understanding. Elevated sensory stimulation ($S(t)$) frequently overwhelms available cognitive resources ($C(t)$), producing a \textit{cognitive tunnel} in which children attend to interface mechanics instead of instructional content. Systems that depend on text-based instructions ($S(t)$) further exclude pre-literate users ($D(t)$), which forces continuous adult mediation. Environmental constraints ($E(t)$) compound these challenges through limited \textbf{teacher readiness}. 

\subsection{The Risk vs. Attention Gap}
We also mapped mean scholarly attention scores against independently derived real-world risk scores for each category, as shown in Figure~\ref{fig:risk-matrix}. This analysis highlights areas where the research environment ($E(t)$) fails to adequately protect child development ($D(t)$), system sensing reliability ($S(t)$), or cognitive workload boundaries ($C(t)$). \textbf{Data security} receives the least scholarly attention (0.14) despite carrying the maximum risk score (9). This imbalance reflects insufficient environmental governance ($E(t)$) for safeguarding children's developmental profiles ($D(t)$) against irreversible biometric data exposure. \textbf{Disability access} shows a similar neglect (Attention: 0.52), indicating that many systems optimize sensing and interaction mechanisms ($S(t)$) for neurotypical users rather than accommodating developmental variability ($D(t)$). \textbf{Privacy} (Attention: 1.04) and \textbf{low-resource access} (Attention: 0.92) both exhibit high risk scores (9). However, existing work frequently treats privacy as a procedural issue managed through parental consent ($E(t)$) instead of addressing technical safeguards such as data minimization and on-device processing ($S(t)$). \textbf{Pedagogy} (Attention: 1.56) and \textbf{technical challenges} (Attention: 0.96) receive the greatest research focus. Studies in these areas emphasize alignment between cognitive demands ($C(t)$) and system stability ($S(t)$) to support learning outcomes. While this focus improves usability, it corresponds to comparatively lower risk levels (4 and 2), since effects such as novelty-driven distraction tend to be transient. \textbf{Health-related issues} (Risk: 4, Attention: 0.81) remain underrepresented and fall within the ignored quadrant. This gap indicates that the field prioritizes cognitive engagement ($C(t)$) while insufficiently addressing the physiological constraints and sensory thresholds of the developing child ($D(t)$).

\section{Conclusion}
The AHD framework provides a diagnostic scaffold that links XR system properties to early childhood developmental constraints. Our analysis of $111$ prior works show that effective XR use in early childhood education requires explicit alignment between cognitive load ($C(t)$), sensory stimulation ($S(t)$), and environmental context ($E(t)$), and the child's developmental profile ($D(t)$). These dimensions jointly determine whether XR supports learning or introduces avoidable friction through overload, sensory conflict, or classroom-level deployment barriers. We therefore advocate a~\textbf{child-centered XR by design} approach. Designers and educators should prioritize security and privacy controls, build for accessibility and adopt interaction vocabularies that accommodate pre-literacy, limited fine motor control, and age-specific sensory thresholds. XR should complement broader learning sequences, using adult mediation to ensure developmentally appropriate demands.

\acknowledgments{
We would like to acknowledge the Data Agency and Security (DAS) Lab at George Mason University and Coventry University where the study was conducted. The opinions expressed
in this work are solely those of the authors.}

\bibliographystyle{abbrv-doi}

\bibliography{IEEEVR2026}

@article{bexson2024safety,
  title={{Safety of virtual reality use in children: a systematic review}},
  author={Bexson, Charlotte and Oldham, Geralyn and Wray, Jo},
  journal={European journal of Pediatrics},
  volume={183},
  number={5},
  pages={2071--2090},
  year={2024},
  publisher={Springer}
}

@article{fast2018testing,
  title={Testing and validating Extended Reality (xR) technologies in manufacturing},
  author={Fast-Berglund, {\AA}sa and Gong, Liang and Li, Dan},
  journal={Procedia Manufacturing},
  volume={25},
  pages={31--38},
  year={2018},
  publisher={Elsevier}
}

@misc{unesco_ecce_need_know_2025,
  author={UNESCO},
  title={What you need to know about early childhood care and education},
  year={2025},
  url={https://www.unesco.org/en/early-childhood-education/need-know},
  note= {Last updated 13 February 2025}
}

@article{noah2021exploring,
  title={Exploring evolution of augmented and virtual reality education space in 2020 through systematic literature review},
  author={Noah, Naheem and Das, Sanchari},
  journal={Computer Animation and Virtual Worlds},
  volume={32},
  number={3-4},
  pages={e2020},
  year={2021},
  publisher={Wiley Online Library}
}
\end{document}